# PepMLM: Target Sequence-Conditioned Generation of Therapeutic Peptide Binders via Span Masked Language Modeling

Tianlai Chen,[1] Madeleine Dumas,[2,3] Rio Watson,[1] Sophia Vincoff,[1] Christina Peng,[4]
Lin Zhao,[1] Lauren Hong,[1] Sarah Pertsemlidis,[1] Mayumi Shaepers-Cheu,[2] Tian Zi Wang,[1] Divya Srijay,[1]
Connor Monticello,[4] Pranay Vure,[1] Rishab Pulugurta,[1] Kseniia Kholina,[1] Shrey Goel,[1]
Matthew P. DeLisa,[5-7] Ray Truant,[4] Hector C. Aguilar,[2] Pranam Chatterjee[1,8,9,†]

1. Department of Biomedical Engineering, Duke University
2. Department of Microbiology and Immunology, College of Veterinary Medicine, Cornell University
3. Department of Microbiology, College of Agriculture and Life Sciences, Cornell University
4. Department of Biochemistry and Biomedical Sciences, McMaster University
5. Meinig School of Biomedical Engineering, Cornell University, Ithaca, NY, USA
6. Robert F. Smith School of Chemical and Biomolecular Engineering, Cornell University, Ithaca, NY, USA
7. Cornell Institute of Biotechnology, Cornell University, Ithaca, NY, USA
8. Department of Computer Science, Duke University
9. Department of Biostatistics and Bioinformatics, Duke University

[†]Corresponding author: pranam.chatterjee@duke.edu

## Abstract

Target proteins that lack accessible binding pockets and conformational stability have posed increasing challenges for drug development. Induced proximity strategies, such as PROTACs and molecular glues, have thus gained attention as pharmacological alternatives, but still require small molecule docking at binding pockets for targeted protein degradation. The computational design of protein-based binders presents unique opportunities to access "undruggable" targets, but have often relied on stable 3D structures or structure-influenced latent spaces for effective binder generation. In this work, we introduce **PepMLM**, a target sequence-conditioned generator of *de novo* linear peptide binders. By employing a novel span masking strategy that uniquely positions cognate peptide sequences at the C-terminus of target protein sequences, PepMLM fine-tunes the state-of-the-art ESM-2 pLM to fully reconstruct the binder region, achieving low perplexities matching or improving upon validated peptide-protein sequence pairs. After successful *in silico* benchmarking with AlphaFold-Multimer, outperforming RFDiffusion on structured targets, we experimentally verify PepMLM's efficacy via fusion of model-derived peptides to E3 ubiquitin ligase domains, demonstrating endogenous degradation of emergent viral phosphoproteins and Huntington's disease-driving proteins. In total, PepMLM enables the generative design of candidate binders to any target protein, without the requirement of target structure, empowering downstream therapeutic applications.

## Introduction

The development of therapeutics largely relies on the ability to design small molecule- or protein-based binders to pathogenic target proteins of interest.[1] These binders can either be used as inhibitors or as functional recruiters of effector enzymes.[2] For example, proteolysis targeting chimeras (PROTACs) or molecular glues are heterobifunctional small molecules that bind and recruit endogenous E3 ubiquitin ligases for targeted protein degradation (TPD).[3,4] Still, these small molecule-based methods rely on the existence of accessible cryptic or canonical binding sites, which are not present on classically "undruggable" intracellular proteins.[5,6] With the advent of deep learning-based structure prediction tools such as AlphaFold2 and AlphaFold3,[7,8] combined with generative modeling,[1] algorithms such as RFDiffusion and MASIF-Seed enable researchers to conduct *de novo* protein binder design from target structure alone.[9,10] Nonetheless, much of the undruggable proteome, including dysregulated proteins such as transcription factors and fusion oncoproteins, are conformationally disordered, thus biasing design to a small subset of disease-related proteins.[1,6]

Over the past few years, deep learning has revolutionized natural language processing (NLP), particularly through the implementation of the attention mechanism.[11] This foundational advancement has transcended the boundaries of natural language analysis, finding applications in the modeling of other languages, such as proteins, which are fundamentally sequences of amino acids.[12] In recent times, several protein language models (pLMs), trained on distinct transformer architectures, such as ProtT5, ProGen2, ProtGPT2, and the ESM series, have accurately captured critical physicochemical properties of proteins.[13–16] Notably, ESM-2 currently stands as the state-of-the-art model in the realm of protein sequence encoding, essentially functioning as an encoder-only model that discerns co-evolutionary patterns among protein sequences via a masked language modeling (MLM) training task.[17,18] These models have been extended to powerful applications, including antibody design, the creation of novel proteins, and structure prediction, offering a streamlined approach to embedding useful protein information.[14,15,17,18] Recently, our lab has leveraged the expressivity of pLMs to both generate and prioritize effective peptidic binder motifs to targets of interest, enabling design of peptide-guided protein degraders[19–21] that are modeled after the ubiquibody architecture developed by Portnoff, et al.[22] As such, uAbs now represent a programmable, CRISPR-like approach for TPD. Our early models, Cut&CLIP and SaLT&PepPr, rely on the existence of interacting partner sequences as scaffolds for peptide design.[20,21] Most recently, our PepPrCLIP model generates *de novo* peptides by first sampling the ESM-2 latent space for naturalistic peptide candidates, and then screening these candidates through a contrastive model to determine target sequence specificity.[23] However, a purely *de novo*, target sequence-conditioned binder design algorithm has yet to be developed.

To achieve this goal, we introduce **PepMLM**, a novel **P**eptide binder design algorithm via **M**asked **L**anguage **M**odeling, built upon the foundations of ESM-2.[17] PepMLM innovates by employing a span masking strategy that uniquely positions the entire peptide binder sequence at the terminus of target protein sequences, compelling ESM-2 to reconstruct the entire binding region (Figure 1A). PepMLM-derived linear peptides achieve low perplexities, matching or improving upon validated peptide-protein sequence pairs in the test dataset, outperform the state-of-the-art RFDiffusion model for peptide generation on structured targets *in silico*,[9] and experimentally exhibit degradation of difficult-to-drug drivers of Huntington's disease and emergent viral phosphoproteins when incorporated into the uAb architecture. Overall, by focusing on the complete reconstruction of peptide regions, PepMLM represents the first example of target-conditioned *de novo* binder design from sequence alone, thus facilitating a deeper understanding of binding dynamics and paving the way for the development of more effective, therapeutic binders to conformationally diverse proteins of interest.

## Results

**PepMLM leverages span masking on ESM-2 embeddings for *de novo* generation of target-binding peptides**

We trained PepMLM using existing peptide-protein binding data sourced from the recent PepNN training set and the gold-standard Propedia dataset.[24,25] We subjected our curated dataset to a filtration process based on the lengths of the binder and target protein sequences, which were confined to 50 and 500 respectively. To remove redundancies, we applied a homology filter thresholded at 80%, resulting in a final training set of

10,000 samples and leaving 203 samples for testing.[26] Each entry in the dataset comprised a concatenated protein and binder sequence. During the training phase, we masked the entire peptide sequence, tasking the model to reconstruct them via the ESM-2-650M model. The discrepancy between the ground truth binder and the reconstructed binder induces a cross entropy loss, thereby forcing parameter updates via gradient descent. Post fine-tuning, we generate peptide binders of specific lengths by providing the model with a target protein sequence and a user-defined number of mask tokens, as illustrated in Figure 1A. Final settings and hyperparameters used to train our model are presented in Supplementary Table 1.

We considered two distinct decoding strategies during the generation phase. The default strategy, akin to ESM-2 or BERT-style models,[17] employs greedy decoding, wherein the token with the highest probability is selected at each site. Despite its efficacy, greedy decoding is limited to the generation of a single peptide binder. To augment the diversity of the peptides, we introduced top-$k$ sampling, allowing PepMLM to randomly select from the top $k$ probable tokens at each site. In this decoding strategy, we evaluated perplexity alongside various $k$ values, ranging from 2 to 10, on the test set of target proteins. For each target protein, we generated 10 binders of the same length as the ground truth binder. We observed that as $k$ increased, perplexity also rose, indicating a decrease in model confidence (Supplementary Figure 1). While higher $k$ values yielded more diverse binder sequences, they also corresponded with an increase in the number of outliers. To find a balance between sequence diversity and maintaining model confidence (as indicated by the lower perplexity), we settled on $k$ = 3 as our final selection.

To substantiate the efficacy of the generated peptides, we conducted a comprehensive series of computational benchmarks with test set peptide-target pairs. The total 203 test set target proteins were utilized to generate one peptide binder each, employing pre-trained ESM-2 embeddings and PepMLM. Subsequently, the pseudo-perplexity of the binder region was computed for four groups of target protein:binder pairs. For a majority of the test set, known binders exhibited a reasonable perplexity range, with only a few outliers (those with a perplexity > 40), validating the PepMLM's effective ability to model them accurately (Figure 1B and Supplementary Table 2). A comparative analysis revealed that the binders generated by PepMLM exhibited lower perplexity values, suggesting a higher likelihood of them making stable binding interactions with the target (Figure 1B). Moreover, our distribution analysis revealed that PepMLM closely mirrors the distribution peak of real binders, a deviation from the distribution shifts observed with the original ESM-2 model alone and with randomly generated binders (Figure 1C). We co-folded two top generated binders, exhibiting high ipTM scores, with their respective target proteins using AlphaFold-Multimer through ColabFold,[27] and overlayed their positions with that of PDB-validated test binders to those targets (Figure 1D). We observe high alignment between the generated and test peptides, highlighting the model's proficiency in capturing the inherent conditional distributions associated with peptide-protein binding.

**PepMLM performs strongly in comparison to RFDiffusion on structured targets *in silico* and *in vitro***

Next, to benchmark PepMLM's generation quality, we co-folded the test and generated binders with their respective target proteins utilizing AlphaFold-Multimer, which has been proven effective at predicting peptide-protein complexes.[28,29] The pLDDT and ipTM scores, verified metrics within AlphaFold2,[7] function as critical indicators of the structural integrity and the potential interface binding affinity of peptide-protein complex, respectively, providing a quantitative assessment of our generation. The extracted ipTM and pLDDT values from our benchmarking indicated a statistically significant negative correlation ($p<0.01$) with PepMLM perplexity, affirming the model's reliability at prioritizing binders with stable binding capacity to the target (Supplementary Figure 2). Subsequent analysis involved sorting the test set based on their ipTM values and contrasting these with the ipTM values of the associated PepMLM-generated binders. Our analysis yielded a hit rate exceeding 38% (Figure 2A). When applying the same evaluation process to RFDiffusion for binder design on the test set, the hit rate was below 30% (Figure 2B), suggesting PepMLM's comparative advantage in designing peptide binders to structured targets, potentially reducing the need for extensive downstream experimental screening.

When evaluating generated peptide binders with ipTM scores surpassing those of the test binders, we classified them into three distinct groups based on ipTM score thresholds: Class I (both test and generated binders with ipTM ≥ 0.7), Class II (generated binders with ipTM ≥ 0.7, but test binders with ipTM < 0.7), and Class III (both generated and test binders with ipTM ≤ 0.7). For each class, three representative complexes were chosen for joint visualization with the test binder (Supplementary Figure 3 and Supplementary Table 3). Observations from Classes I and II indicate that despite the generated binders possessing distinct sequence compositions compared to the test binders, they tend to target the same binding pocket and exhibit similar structural conformations. This pattern suggests that our language model-based design approach successfully captures structural information of peptide-protein binding. Conversely, in Class III, characterized by lower ipTM values, we noted distinct binding modes between generated and test binders. The generated binders appeared to occupy more optimal binding positions according to AlphaFold-Multimer predictions (Supplementary Figure 3). However, even with the high pLDDT values from AlphaFold, it remains challenging to definitively ascertain whether our binders exhibit unique binding modes or if these observations are attributable to limitations in AlphaFold2-Multimer modeling. The recent AlphaFold3 model, once open-sourced, may provide stronger predictions in future structure-based benchmarking efforts.[8]

To overcome these shortcomings, we sought to experimentally test PepMLM vs. RFDiffusion-generated binders to the extracellular domain (ECD) of NCAM1/CD56, a protein involved in cell-to-cell adhesion, which plays critical roles in neural development and synaptic plasticity, and has been implicated in cancer progression.[30] Using the ECD sequence of NCAM1 for PepMLM input and its structure for RFDiffusion input, we generated and synthesized four linear peptides from each method (Supplementary Table 4) and performed a sandwich ELISA at an NCAM1 ECD concentration of 1.37 µM, using the peptide as the capture binder. We first observed a significant increase in absorbance between each peptide's interaction with NCAM1 vs. the negative PBS control. We further noted that the top PepMLM peptide, NCAM1_pMLM_4, demonstrates significantly stronger binding than all four RFDiffusion peptides at the equivalent concentration of NCAM1 (Figure 2C).

**PepMLM-derived uAbs degrade Huntington's disease-related proteins *in vitro***

Having demonstrated PepMLM's comparatively strong binder generation to RFDiffusion in both *in silico* and *in vitro* contexts, we next evaluated PepMLM peptides via fusion to E3 ubiquitin ligase domains, generating uAbs to degrade pathogenic proteins in human cells (Figure 3A).[31] We focused our attention on Huntington's disease, a monogenic dominant neurological disorder affecting more than 1 in 10000 adults, caused primarily by an expanded CAG repeat in exon 1 of the *HTT* gene, thus producing an extended polyglutamine (polyQ) tract and resulting in aggregation-prone mutant *huntingtin* protein (mHTT).[32] Recently, it has been shown that genetic knockdown of the mismatch repair-associated MSH3 protein reduces and inhibits mHTT repeat expansion.[33,34] Here, we thus sought to degrade MSH3 at the post-translational level.

First, to design peptides for MSH3 degradation, we employed greedy decoding to determine the optimal binder length that yielded the lowest perplexity, followed by the generation of binders using top $k$ sampling, where $k$ was fixed at 3 as previously described (Supplementary Table 4). After cloning these peptides into our uAb backbone and transfecting into HEK293T cells, which express MSH3 at high levels, we conducted Western blotting on whole-cell protein extracts with MSH3-selective primary antibodies. Our results demonstrate that select PepMLM-generated "guide" peptides, most notably MSH3_pMLM_7, induce robust degradation of MSH3 when fused to E3 ubiquitin ligase domains, demonstrating reduced protein levels relative to that of the non-targeting control poly-glutamine uAb (Figure 3B and Supplementary Table 4). We next sought to degrade the mHTT protein itself. To do this, we utilized TruHD fibroblasts, a genomically stable line which expresses the mHTT protein at a clinically-relevant CAG repeat length of Q43.[35] As the line is heterozygous with both Q43/Q17 alleles, we designed PepMLM peptides targeting exon 1 with a polyQ repeat of 43, and screened these peptides for those with high perplexities for, and thus poor binding prediction to, the Q17 variant. After down-selecting 5 optimal candidates (Supplementary Table 4), we transfected the TruHD line and measured HTT degradation with the EPR5526 antibody recognizing the first 100 amino acids of HTT that includes the polyQ region. We show that 3 of the 5 PepMLM-designed peptides demonstrate robust degradation of HTT, with Q43_pMLM_3 demonstrating almost complete ablation (Figure 3C). Future work will establish the specificity of these peptides to the Q43 variant.

**PepMLM-derived uAbs degrade emergent viral phosphoproteins**

Finally, we investigated whether PepMLM-derived uAbs could induce degradation of critical viral target proteins. As a key target class, we selected the viral phosphoprotein (P) based on its relatively high sequence homology amongst strains of the selected viruses, as well as its critical role in viral transcription and genome replication. P sequences were selected for two emerging deadly viruses with high pandemic potential, the henipaviruses Nipah virus (NiV) and Hendra virus (HeV), both of which pose significant threats to human health with recorded mortality rates of 50-100%.[36,37] A third P sequence was selected for the endemic virus human metapneumovirus (HMPV), whose infections occur more frequently than NiV and HeV, displaying seasonal cold-like symptoms that are severe and sometimes fatal in young children and elderly populations.[36] There are few to no vaccines or antiviral treatments approved for human use for these three viruses. For these three P proteins, 60 PepMLM-designed uAbs were designed (Supplementary Table 4) and screened for their ability to induce proteasomal degradation of their respective viral P target through Western blot analysis (Figure 4). Of the 60 uAbs screened, we observed a total of 16 demonstrating 25-49% reduction in phosphoprotein level and 8 with over 50% P protein reduction, suggesting an overall hit-rate of around 40%, in strong agreement with our *in silico* hit rate shown in Figure 2A (Figure 4A-C). Specifically, seven HMPV-targeting uAbs exhibited over 50% reduction of HMPV P protein expression and uAbs NiV_2, 3, 8, 11, 12, and 17 demonstrated moderate to strong degradation of the Nipah virus P protein. HeV_2 showed the most drastic change in henipaviral P protein presence with a ~56% reduced detection of HeV P. Together, these results provide strong evidence of PepMLM's ability to accelerate development of therapeutics to current and emergent diseases.

## **Discussion**

By simply redesigning a guide RNA, the CRISPR-Cas system enables targeting and modification of almost any DNA sequence, a programmable method that has revolutionized biology.[38] Specifically, with the recent engineering of protospacer adjacent motif (PAM)-relaxed Cas variants, there is minimal restriction as to which user-defined DNA sequences can be bound and edited.[39,40] To enable similar programmable targeting of any protein, here, we introduce PepMLM, the first *de novo* binder design algorithm directly conditioned only on the target sequence of a protein, without any structural requirement. By using generated peptides as guides for E3 ubiquitin ligase domains, and eventually other post-translational modification domains, this work serves as a step forward towards developing a fully modular proteome editing system.

To this point, we had utilized the lightweight ESM-2-650M model, enabling flexible fine-tuning and inference. To assess the performance of larger models, we note that we additionally fine-tuned ESM-2-3B[17] for peptide generation (PepMLM-3B) and evaluated it using the same methodology as employed for the ESM-2-650M version of PepMLM (PepMLM-650M). However, as illustrated in Supplementary Figure 4, we did not observe a substantial improvement in either perplexity or hit rate for PepMLM-3B (36.02%). Considering the associated resource and inference costs, we provide our PepMLM-650M model as an accessible resource for effective linear peptide generation.

Nonetheless, we envision that further improvements can be made to PepMLM-650M, enabling its adoption as a universal tool for peptide binder design. For example, PepMLM can be retrained with modification-aware and variant-aware pLM embeddings to enable specificity to post-translational isoforms over wild-type protein states.[41] Our future experimental work directions will include biochemical and molecular validation and characterization of the antiviral therapeutic potential of top selected uAbs within the ME and CE groups tested for the emergent viral targets. We also plan to integrate PepMLM generation with high-throughput lentiviral screening to further evaluate its hit rate and input experimental data back into the algorithm, creating an active learning-based optimization loop. As a note, we have not applied any experimental optimization of PepMLM-derived peptide binders, including further stabilization via cyclization or stapling, modifications which may improve therapeutic utility.[42,43] In total, we envision that through additional development, our accessible peptide generator, coupled with variants of our uAb architecture, will enable a CRISPR-analogous system to bind and modulate any target protein, whether structured or not.

## Methods

**Data Curation**

In the data curation phase, protein and peptide complexes were amalgamated from the PepNN and Propedia databases.[24,25] Initially, redundancy between the two datasets was eliminated, followed by the utilization of MMseqs2 to cluster the remaining protein sequences, setting a threshold of 0.8.[26] When protein sequences were identified within the same cluster and exhibited identical binder sequences, a single sequence was retained. This was followed by a manual filtering process, wherein protein sequences were sorted and those exhibiting high similarity (threshold of 80%) were removed to further mitigate homology issues. Consequently, a dataset comprising 10,203 entries was amassed, from which 10,000 were randomly allocated for training and 203 for testing. The maximum lengths for the binder and protein sequences were established at 50 and 500, respectively.

**Conditional Peptide Modeling**

Peptide binders are modeled in a distinctive manner, wherein the peptides are modeled conditionally based on the full protein sequence. Let $p = (p_1, p_2, p_3, \ldots, p_n)$ represent the target protein sequence of length $n$ and $b = (b_1, b_2, b_3, \ldots, b_m)$ denote the binder of length $m$. The protein and peptide sequences are concatenated, incorporating special tokens of start, end, and padding. Mask language modeling transforms this into a conditional modeling problem, where the objective is to reconstruct $b$ given $p$ and entire masked $b$ region. In natural language processing, such a technique is close to span masking which has demonstrated enhanced performance in language modeling.[44,45] The entire model is updated with Masked Language Model (MLM) loss, which can be represented as:

$$\mathcal{L}_{\mathrm{MLM}} = -\tfrac{1}{m} \sum_{i \in m} \log P(b_i | p, b_{\mathrm{mask}})$$

Through this methodology, the discrepancy between the generated binders and the ground truth is minimized, facilitating the approximation of the conditional probability, $\prod_{i=1}^{m} P(b_i|p)$.

**PepMLM Training**

The pre-trained protein language model, ESM-2, was utilized to facilitate full parameter fine-tuning. ESM-2, a transformer-based model, is adept at discerning coevolutionary patterns across protein sequences. The concatenated protein and peptide sequences were tokenized at the amino acid level and input into the model. Deviating from the original training strategy of ESM, the entire binder sequence was exclusively masked, compelling the model to learn the relationship between the peptide binder and the protein. The ESM-2-650M and ESM-2-3B models were both trained for PepMLM. Both versions were trained on an NVIDIA 8xA100 640 GB DGX GPU system with Pytorch 2.01 and Python 3.10.10. Specific parameters are shown in Supplementary Table 1.

**PepMLM Generation**

During the generation phase, the target protein sequence, along with a designated number of mask tokens (at end), was input into the model. Subsequently, the model greedily decodes logits at each masked position to identify peptide binders. To infuse greater diversity into the generation process, top *k* sampling was implemented, wherein the model randomly selects the top *k* highest probability logits at each masked position.

**Pseudo-Perplexity of PepMLM**

The pseudo-perplexity of ESM-2 was adapted to focus specifically on the evaluation of peptide binder generation. Notably, the perplexity calculation is confined to the binder region, or, in other words, the masked regions. Mathematically, the pseudo-perplexity is defined as:

$$\mathrm{PseudoPerplexity}(b) = \exp\left\{-\tfrac{1}{m} \sum_{i=1}^{m} \log p\left(b_i | b_{j \neq i}, p\right)\right\}$$

In this equation, $b$ represents the binder sequence and $m$ is the length of the binder sequence. This modification ensures a more focused evaluation of the generated peptide binders, aligning with the conditional modeling approach adopted in this study.

**Generated Peptide Benchmarking**

To assess the efficacy of the generated peptide binders, two benchmarking studies were conducted: one on the test set and another on selected critical proteins. In the test set benchmarking, top-*k* sampling (*k* = 3) was employed to generate a single peptide binder for each target protein. Additionally, the original ESM-2 model was utilized to generate peptides, and random peptides of equivalent length were created. For ESM-2 generation, specifically, mask tokens of the same length were added at the end of target protein sequences for analogous model prediction and decoding as for PepMLM. The perplexity of the PepMLM was compared across four groups. PepMLM-generated binders and test binders were folded using the AlphaFold2 ColabFold version 1.5.2, in conjunction with the protein sequences. Folding metrics including pLDDT and ipTM were gathered, which were utilized to correlate perplexity findings. For each test target protein, the ipTM scores of the test and generated binders were compared to determine the overall hit rate. Notice, as top-*k* sampling generates with randomness, the hit rate might vary or increase with different runs or *k* options. For the proteins identified as critical, the model produced eight binders, each of a length of 15 residues, using top-*k* sampling (*k* = 3). These binders were synthesized for specific target proteins to facilitate subsequent experimental evaluations.

**Co-Folding Complex Visualization**

For the visualization of AlphaFold-Multimer co-folding results from PepMLM-generated binder-protein complexes, an initial alignment with the corresponding test complex was performed using Biopython version 1.8.3, facilitated a comparative visualization of selected complexes, encompassing both the generated and test binders. In these visualizations, the target protein was depicted in yellow, contrasting with the test and generated binders, colored in blue and red, respectively. The visualizations were executed using py3Dmol version 2.0.4.

**RFDiffusion Generation**

In parallel to the PepMLM approach, RFDiffusion was employed to design peptide binders for both cases. For the given test set, RFDiffusion was tasked with generating one peptide binder per target protein, matching the length specified by the ground truth binders. The predicted structures were then converted into sequences using ProteinMPNN with initial guess and number of cycles of 3. For the selected critical proteins, RFDiffusion and ProteinMPNN generated 8 candidate binders, each comprising 15 residues, under identical parameter settings as testset generation. RFDiffusion inference code on ColabFold can be found here: https://colab.research.google.com/github/sokrypton/ColabDesign/blob/v1.1.1/rf/examples/diffusion.ipynb

**Sandwich ELISA**

Linear peptides to NCAM1 (Supplementary Table 4) were purchased at >80% purity from CPC Scientific. In brief, 96-well plates (Corning, 9018) were coated with 2 µg/mL of peptides (Biomatik RPU40704) diluted in 1xPBS, pH 7.4, at a volume of 50 µL per well in 4 °C overnight. Following this, the plates were washed with 1x PBST (PBS, 0.1% (v/v) Tween 20) three times and added 200 µL of blocking buffer (3% bovine serum albumin (BSA) in 1xTBST) per well overnight at 4 °C. PBS or recombinant NCAM1 (Manufacturer Product No.) were diluted in triplicate at 1.37 µM in 1xPBS and added to the ELISA plates for 1 h at 37 °C. The plates were washed three times with 1xPBS-T, then samples were treated with biotinylated Anti-CD56 (NCAM) monoclonal antibody (ThermoFisher, Cat # 13-0567-82; diluted 1:5000). The plates were then washed three times with 1xPBS-T and incubated for 1 h at room temperature with HRP-conjugated streptavidin (ThermoFisher, Cat # N100; diluted 1:20,000), with shaking at 450 rpm. Following an additional three washes with 1xPBS-T, 100 µL of 3,3'-5,5'-tetramethylbenzidine substrate (1-Step Ultra TMB-ELISA; ThermoFisher, 34029) was added to each well, and the plates were incubated at room temperature in darkness. The reaction was quenched by adding 100 µL of 2 M $H_2SO4$, and absorbance was quantified at a wavelength of 450 nm utilizing a FilterMax F5 microplate spectrophotometer (Agilent).

**Generation of plasmids**

All uAb plasmids were generated from the standard pcDNA3 vector, harboring a cytomegalovirus (CMV) promoter and a C-terminal P2A-GFP cassette as a transfection control. An Esp3I restriction site was introduced immediately upstream of the CHIPΔTPR CDS and flexible GSGSG linker via the KLD Enzyme Mix (NEB) following PCR amplification with mutagenic primers (Genewiz). For uAb assembly, PepMLM-derived peptide sequences (Supplementary Table 4) were human codon-optimized for complementary oligo generation (Genewiz). Oligos were annealed and ligated via T4 DNA Ligase into the Esp3I-digested uAb backbone. Assembled constructs were transformed into 50 µL NEB Turbo Competent *Escherichia coli* cells, and plated onto LB agar supplemented with the appropriate antibiotic for subsequent sequence verification of colonies and plasmid purification (Genewiz).

Sequences for human codon-optimized phosphoprotein genes for NiV (GenBank AY029767), HeV (GenBank MN062017), and HMPV (GenBank AAS22075) were designed with HA tags on their N-termini and flanked with restriction enzyme recognition sites for KpnI and XhoI on their 3' and 5' ends, respectively, for ligation into a mammalian pCAGGS vector.

**Cell culture**

HEK293T cells were maintained in Dulbecco's Modified Eagle's Medium (DMEM) supplemented with 100 units/ml penicillin, 100 mg/ml streptomycin, and 10% fetal bovine serum (FBS). The PepMLM peptides (500 ng) plasmids were transfected into cells (4x10$^5$/well in a 12-well plate) with Lipofectamine 2000 (Invitrogen) in Opti-MEM (Gibco). TruHD-Q43Q17M cells were maintained in Minimum Essential Medium Eagle with Earle's Salts (EMEM) supplemented with 15% FBS, 1% NEAA (Gibco), and 1% GlutaMAX (Gibco). The PepMLM peptides were transfected into the fibroblasts using the SG cell line 4D-Nucleofector X kit (Lonza). For viral protein degradation, transfections were done with HEK293T cells at approximately 90% confluency in 6-well plates using a 4:1 µL/µg ratio of polyethylenimine (PEI) Max to DNA, following the transfection reagent manufacturer's protocol. Target P plasmids were transfected at a 1:1 ratio with uAb plasmids for a total of 2 µg of DNA per well in OptiMem. Transfections were supplemented with OptiMem at approximately 5 h post transfection.

**MSH3 Western blotting**

On the day of harvest, cells were washed with 1X PBS and detached by addition of 0.05% trypsin-EDTA, then pelleted by centrifugation at 1000 rcf for 5 min. Cells were then lysed and subcellular fractions were isolated from lysates using a 1:100 dilution of protease inhibitor cocktail (Millipore Sigma) in Pierce RIPA buffer (ThermoFisher). Specifically, the protease inhibitor cocktail-RIPA buffer solution was added to the cell pellet, the mixture was placed on ice for 30 min followed by centrifugation at 15,000 rpm for 15 min at 4 °C. The supernatant was collected immediately to a pre-chilled PCR tube, and the protein concentration was quantified using a Pierce BCA Protein Assay kit (Thermofisher). After adding 4X Bolt™ LDS Sample Buffer (ThermoFisher) with 5% β-mercaptoethanol in a 3:1 ratio to 20µg of protein, the mixture was incubated at 95 °C for 10 min prior to immunoblotting. Immunoblotting was performed according to standard protocols. Briefly, samples were loaded at equal volumes into Bolt™ Bis-Tris Plus Mini Protein Gels (ThermoFisher) and separated by electrophoresis. iBlot™ 2 Transfer Stacks (Invitrogen) were used for membrane blot transfer, and after blocking at 1 h room-temperature in 1% Nonfat Dry Milk (Thermofisher) in 1X TBS-T (Thermofisher), proteins were probed with mouse anti-MSH3 antibody (Santa Cruz Biotechnology, Cat #sc-271080; diluted 1:2000) and mouse anti-GAPDH antibody (Santa Cruz Biotechnology, Cat #sc-47724; diluted 1:10000) for overnight incubation at 4°C. The blots were washed three times with 1X TBST for 10 min each and then probed with a secondary antibody, goat anti-mouse IgG (H+L) Poly-HRP (ThermoFisher, Cat #32230, diluted 1:5000) for 1 h at room temperature. Following three washes with 1X TBST for 10 min each, blots were detected by chemiluminescence using ChemiDoc™ Touch Imaging System (BioRad).

**HTT Western blotting**

For TruHD-Q43Q17M cells, on the day of harvest, cells were washed with 1x PBS, then lysed and scraped off using radioimmunoprecipitation assay buffer (RIPA; 50 mM Tris-HCl pH 8.0, 150 mM NaCl, 1% NP-40, 0.25% sodium deoxycholate, 1 mM EDTA) with protease and phosphatase inhibitors (Thermo) on ice. The mixture was incubated on ice for 5 minutes followed by centrifugation at 13,000 rpm for 5 minutes at 4°C. The supernatant was collected and quantified using the BCA Protein Assay Kit (Sigma). 4x loading buffer (250mM Tris pH 6.8, 40% glycerol, 8% SDS, 0.02% bromophenol blue) was added to the supernatant and incubated at 95°C for 5 minutes. Immunoblotting was performed using precast 4-20% gradient gels (Bio-Rad) and then transferred onto an Immobilon®-P PVDF membrane (Millipore). The membranes were blocked in 5% skim milk powder in 1x TBS-T (50 mM Tris-HCl, pH 7.5,150 mM NaCl, 0.1% Tween-20) at 4°C overnight, then probed with rabbit anti-huntingtin antibody (Abcam, EPR5526, 1:5000) or rabbit anti-vinculin antibody (Abcam, EPR8185, 1:5000) in the same buffer for one hour at room temperature. The membranes were washed three times with 1x TBS-T, then three times with 2.5% skim milk powder in 1x TBS-T for 5 minutes each. The membranes are then probed with horseradish peroxidase-conjugated secondary antibodies (Abcam, 1:50000) for 30 minutes at room temperature before being washed again and incubated with Immobilon Western Chemiluminescent HRP Substrate (Millipore) and imaged with MicroChemi chemiluminescence detector (DNR Bio-imaging Systems).

### Viral phosphoprotein Western blotting

HEK293T cells were harvested 48 hours post-transfection and lysed using 1X radioimmunoprecipitation (RIPA, Millipore) buffer containing complete protease inhibitor (Sigma). The cells were incubated at 4°C, rocking, for 40 min before being vortexed at 5 min intervals for 20 min. Cell lysate supernatants were collected following centrifugation at 21,000 x g for 30 min at 4°C. To denature samples for sodium dodecyl sulfate-polyacrylamide gel electrophoresis (SDS-PAGE), cell lysates were mixed and incubated with 1.8% SDS containing 5% beta-mercaptoethanol for 10 minutes at 95°C before loading onto a 10% acrylamide-Tris HCl gels. Proteins were separated at 100 V for 2 h and then transferred onto 0.2 μm polyvinylidene difluoride (PVDF) membranes at 0.5 A for 2 h. Membranes were blocked in phosphate buffered saline with 0.2% Tween 20 (PBST) containing 4% bovine serum albumin (BSA) before staining in 1:1,000 dilutions of mouse anti-FLAG (Millipore, CAT: F1804), mouse anti-β-actin (Santa Cruz Biotechnology, CAT: 47778), and rabbit anti-HA (Biolegend, CAT: 923502) primary antibodies. Secondary antibody staining was performed using 1:1,000 dilutions of goat anti-mouse Alexa Fluor 647 and goat anti-rabbit Alexa Fluor 488 secondary antibodies (Invitrogen, CAT: A21236 and A11008, respectively). Blocking, primary and secondary antibody membrane incubations were performed rocking at room temperature for 30 min, 1 h, and 30 min, respectively. Membranes were rinsed with PBST 3 times for 5 min each following antibody staining. All membranes were imaged using a Bio-Rad imager in respective Alexa Fluor channels. Densitometric quantification was performed using ImageLab for phosphoprotein and β-actin bands. Background densities from samples mock-transfected with pCAGGS vector only were subtracted. Then, sample densities were normalized to their respective β-actin signals before normalization to their respective phosphoprotein-only controls. Bar graphs were produced using GraphPad Prism, version 10 and the schematic diagram was made using BioRender.

### Statistical analysis and reproducibility

All data were reported as average values with error bars representing standard deviation (SD). For samples performed in independent biological triplicates (n=3) or above, statistical significance was determined by unpaired $t$ test (*$p$ < 0.05, **$p$ < 0.01; ***$p$ < 0.001; ****$p$ < 0.0001). All graphs were generated using Prism 10 for MacOS version 14.4.1. No data were excluded from the analyses. The experiments were not randomized. The investigators were not blinded to allocation during experiments and outcome assessment.

### Author Contributions

T.C. designed the PepMLM architecture, curated peptide-protein data, trained and evaluated trained models. T.C., R.P., and P.V. performed *in silico* benchmarking. L.H. and S.P. conducted sandwich ELISA assays and analyzed results. R.W. performed MSH3 degradation assays, assisted by L.H., D.S., and T.Z.W. L.Z. constructed mHTT uAbs and C.P. performed degradation assays, respectively. M.D. and M.S.C. performed viral phosphoprotein degradation assays. C.M. produced P protein-targeting uAbs. S.V. computationally designed PepMLM peptides for experimental testing, assisted by K.K. and S.G. M.P.D. supervised uAb construction for P protein degradation. R.T. supervised mHTT degradation assays. H.C.A. supervised P protein degradation assays. P.C., T.C., H.C.A. and M.D. wrote the manuscript with input from all authors. P.C. conceived, designed, supervised, and directed the study.

## Data and Materials Availability

All data needed to evaluate the conclusions in the paper are present in the paper and code repository: https://github.com/programmablebio/pepmlm. PepMLM-650M is also hosted on HuggingFace for academic use only, with an easy-to-use demo for peptide generation: https://huggingface.co/ChatterjeeLab/PepMLM-650M. All raw and processed data (including raw immunoblots) have been deposited to the Zenodo repository: https://doi.org/10.5281/zenodo.11201091.

## Competing Interests

P.C. and M.P.D. is a co-founder of UbiquiTx, Inc., and are inventors of patents related to genetically-encoded proteome editing technologies. The remaining authors declare no competing interests.

## Acknowledgements

We thank the Duke Compute Cluster and Mark III Systems for providing database and compute resources that have contributed to the research reported within this manuscript. We also thank Suhaas Bhat and Manvitha Ponnapati for curating initial peptide datasets, and Venkata Srikar Kavirayuni and Ashley Hsu for assisting in PepMLM benchmarking.

## Declarations

The research was supported by institutional startup funds to the lab of P.C. from Duke University, as well as funds from the Wallace H. Coulter Foundation, the CHDI Foundation, The Hartwell Foundation and NIH grants 3U54CA231630-01A1S4 and 1R21CA278468-01 to the lab of P.C. The work was also funded by institutional Cornell funds and NIH grant R01 AI109022 to the lab of H.A.C. and by the Krembil Foundation to the lab of R.T.

## References

1. Chen, T., Hong, L., Yudistyra, V., Vincoff, S. & Chatterjee, P. Generative design of therapeutics that bind and modulate protein states. *Curr. Opin. Biomed. Eng.* **28**, 100496 (2023).
2. Zhong, L. *et al.* Small molecules in targeted cancer therapy: advances, challenges, and future perspectives. *Signal Transduction and Targeted Therapy* **6**, 1–48 (2021).
3. Békés, M., Langley, D. R. & Crews, C. M. PROTAC targeted protein degraders: the past is prologue. *Nat. Rev. Drug Discov.* **21**, 181–200 (2022).
4. Dong, G., Ding, Y., He, S. & Sheng, C. Molecular Glues for Targeted Protein Degradation: From Serendipity to Rational Discovery. *J. Med. Chem.* **64**, 10606–10620 (2021).
5. Gao, H., Sun, X. & Rao, Y. PROTAC Technology: Opportunities and Challenges. *ACS Med. Chem. Lett.* **11**, 237–240 (2020).
6. Behan, F. M. *et al.* Prioritization of cancer therapeutic targets using CRISPR-Cas9 screens. *Nature* **568**, 511–516 (2019).
7. Jumper, J. *et al.* Highly accurate protein structure prediction with AlphaFold. *Nature* **596**, 583–589 (2021).
8. Abramson, J. *et al.* Accurate structure prediction of biomolecular interactions with AlphaFold 3. *Nature* 1–3 (2024).
9. Watson, J. L. *et al.* De novo design of protein structure and function with RFdiffusion. *Nature* **620**, 1089–1100 (2023).
10. Gainza, P. *et al.* De novo design of protein interactions with learned surface fingerprints. *Nature* **617**, 176–184 (2023).
11. Vaswani, A. *et al.* Attention is all you need. (2017) doi:10.48550/ARXIV.1706.03762.
12. Ofer, D., Brandes, N. & Linial, M. The language of proteins: NLP, machine learning & protein sequences. *Comput. Struct. Biotechnol. J.* **19**, 1750–1758 (2021).

13. Elnaggar, A. *et al.* ProtTrans: Toward Understanding the Language of Life Through Self-Supervised Learning. *IEEE Trans. Pattern Anal. Mach. Intell.* **44**, 7112–7127 (2022).
14. Madani, A. *et al.* Large language models generate functional protein sequences across diverse families. *Nat. Biotechnol.* **41**, 1099–1106 (2023).
15. Ferruz, N., Schmidt, S. & Höcker, B. ProtGPT2 is a deep unsupervised language model for protein design. *Nat. Commun.* **13**, 1–10 (2022).
16. Rives, A. *et al.* Biological structure and function emerge from scaling unsupervised learning to 250 million protein sequences. *Proc. Natl. Acad. Sci. U. S. A.* **118**, (2021).
17. Lin, Z. *et al.* Evolutionary-scale prediction of atomic-level protein structure with a language model. *Science* **379**, 1123–1130 (2023).
18. Hie, B. L. *et al.* Efficient evolution of human antibodies from general protein language models. *Nat. Biotechnol.* 1–9 (2023).
19. Chatterjee, P. *et al.* Targeted intracellular degradation of SARS-CoV-2 via computationally optimized peptide fusions. *Communications Biology* **3**, 1–8 (2020).
20. Palepu, K. *et al.* Design of Peptide-Based Protein Degraders via Contrastive Deep Learning. *bioRxiv* (2022) doi:10.1101/2022.05.23.493169.
21. Brixi, G. *et al.* SaLT&PepPr is an interface-predicting language model for designing peptide-guided protein degraders. *Communications Biology* **6**, 1–10 (2023).
22. Portnoff, A. D., Stephens, E. A., Varner, J. D. & DeLisa, M. P. Ubiquibodies, synthetic E3 ubiquitin ligases endowed with unnatural substrate specificity for targeted protein silencing. *J. Biol. Chem.* **289**, 7844–7855 (2014).
23. Bhat, S. *et al.* De Novo Generation and Prioritization of Target-Binding Peptide Motifs from Sequence Alone. *bioRxiv* 2023.06.26.546591 (2023) doi:10.1101/2023.06.26.546591.
24. Abdin, O., Nim, S., Wen, H. & Kim, P. M. PepNN: a deep attention model for the identification of peptide binding sites. *Communications Biology* **5**, 1–10 (2022).
25. Martins, P. *et al.* Propedia v2.3: A novel representation approach for the peptide-protein interaction database using graph-based structural signatures. *Front Bioinform* **3**, 1103103 (2023).
26. Steinegger, M. & Söding, J. MMseqs2 enables sensitive protein sequence searching for the analysis of massive data sets. *Nat. Biotechnol.* **35**, 1026–1028 (2017).
27. Mirdita, M. *et al.* ColabFold: making protein folding accessible to all. *Nat. Methods* **19**, 679–682 (2022).
28. Evans, R. *et al.* Protein complex prediction with AlphaFold-Multimer. *bioRxiv* (2021) doi:10.1101/2021.10.04.463034.
29. Johansson-Åkhe, I. & Wallner, B. Improving peptide-protein docking with AlphaFold-Multimer using forced sampling. *Front Bioinform* **2**, 959160 (2022).
30. Sasca, D. *et al.* NCAM1 (CD56) promotes leukemogenesis and confers drug resistance in AML. *Blood* **133**, 2305–2319 (2019).
31. Zöllner, S. K. *et al.* Ewing Sarcoma-Diagnosis, Treatment, Clinical Challenges and Future Perspectives. *J. Clin. Med. Res.* **10**, (2021).
32. Finkbeiner, S. Huntington's Disease. *Cold Spring Harb. Perspect. Biol.* **3**, (2011).
33. Driscoll, R. *et al.* Dose-dependent reduction of somatic expansions but not Htt aggregates by di-valent siRNA-mediated silencing of MSH3 in HdhQ111 mice. *Sci. Rep.* **14**, 1–11 (2024).
34. O'Reilly, D. *et al.* Di-valent siRNA-mediated silencing of MSH3 blocks somatic repeat expansion in mouse models of Huntington's disease. *Mol. Ther.* **31**, 1661–1674 (2023).
35. Hung, C. L.-K. *et al.* A patient-derived cellular model for Huntington's disease reveals phenotypes at clinically relevant CAG lengths. *Mol. Biol. Cell* **29**, 2809–2820 (2018).
36. Gálvez, N. M. S. *et al.* Host Components That Modulate the Disease Caused by hMPV. *Viruses* **13**, (2021).
37. Gazal, S. *et al.* Nipah and Hendra Viruses: Deadly Zoonotic Paramyxoviruses with the Potential to Cause the Next Pandemic. *Pathogens* **11**, (2022).
38. Wang, J. Y. & Doudna, J. A. CRISPR technology: A decade of genome editing is only the beginning. *Science* **379**, eadd8643 (2023).
39. Walton, R. T., Christie, K. A., Whittaker, M. N. & Kleinstiver, B. P. Unconstrained genome targeting with near-PAMless engineered CRISPR-Cas9 variants. *Science* **368**, 290–296 (2020).
40. Zhao, L. *et al.* PAM-flexible genome editing with an engineered chimeric Cas9. *Nat. Commun.* **14**, 1–8 (2023).

41. Peng, Z., Schussheim, B. & Chatterjee, P. PTM-Mamba: A PTM-Aware Protein Language Model with Bidirectional Gated Mamba Blocks. *bioRxiv* 2024.02.28.581983 (2024) doi:10.1101/2024.02.28.581983.
42. Vinogradov, A. A., Yin, Y. & Suga, H. Macrocyclic Peptides as Drug Candidates: Recent Progress and Remaining Challenges. *J. Am. Chem. Soc.* **141**, 4167–4181 (2019).
43. Moiola, M., Memeo, M. G. & Quadrelli, P. Stapled Peptides-A Useful Improvement for Peptide-Based Drugs. *Molecules* **24**, (2019).
44. Joshi, M. *et al.* SpanBERT: Improving pre-training by representing and predicting spans. (2019) doi:10.48550/ARXIV.1907.10529.
45. Raffel, C. *et al.* Exploring the limits of transfer learning with a unified text-to-text transformer. (2019) doi:10.48550/ARXIV.1910.10683.

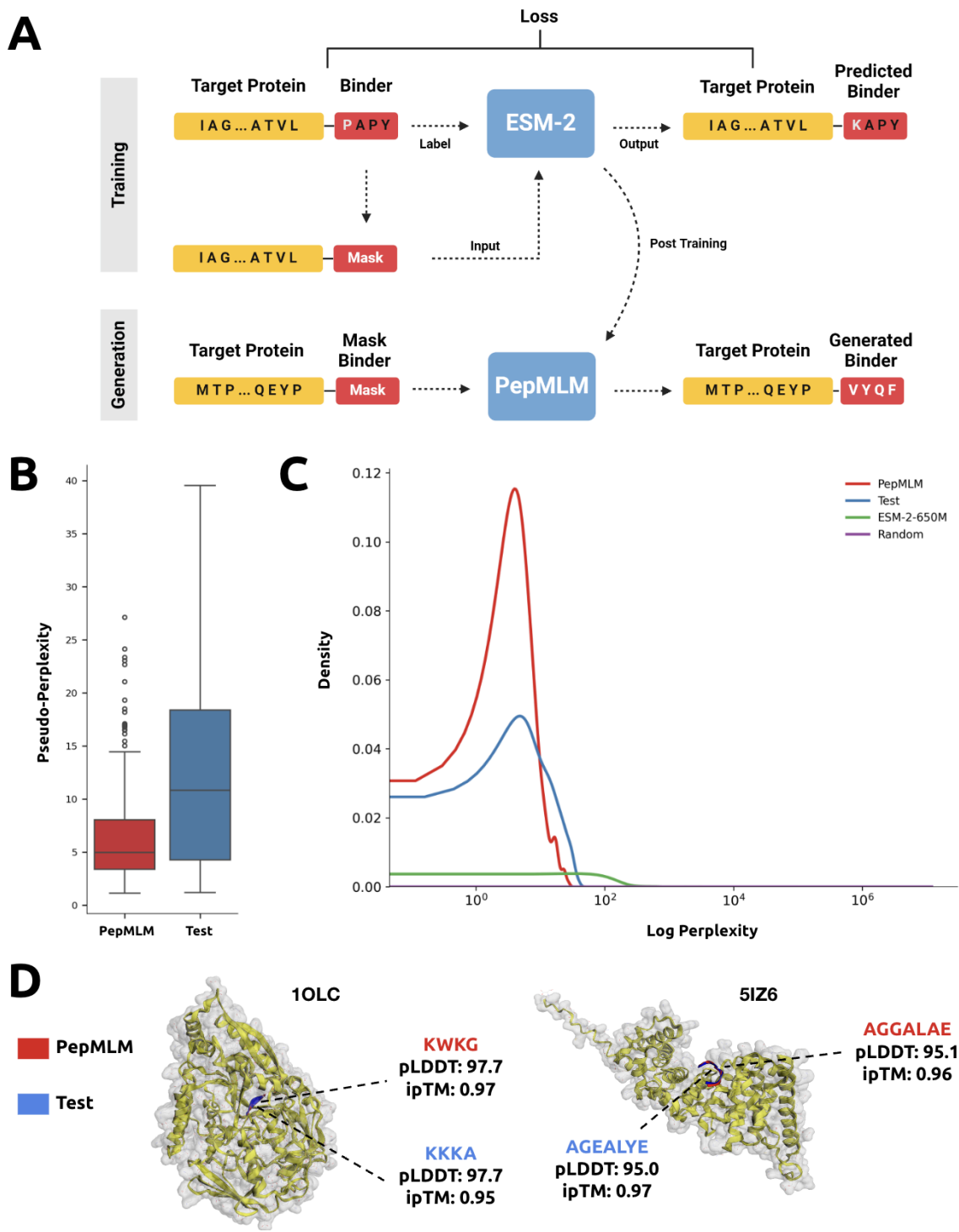

**Figure 1. Overview and evaluation of the PepMLM model.** (A) The architecture of the PepMLM model. Based on the fine-tuning of ESM-2, the model incorporates the target protein sequence along with a masked binder region during the training phase. During the generation phase, the model can accept target protein sequences and mask tokens to facilitate the creation of peptides of specified lengths. (B) Perplexity distribution comparison. The perplexity values were calculated for test and generated peptides, encompassing the target proteins in the test set. (C) The density distribution visualization of the log perplexity values for target-peptide pairs, encompassing test peptides, PepMLM-650M-generated peptides, ESM-2-650M-generated peptides, and random peptides. (D) AlphaFold-Multimer co-folding examples of specified target proteins from the PDB and sampled peptide binders generated via PepMLM-650M, with the pLDDT values serving as the determinant for color coding. ipTM scores indicate stability of the binding interface, a proxy for binding affinity.

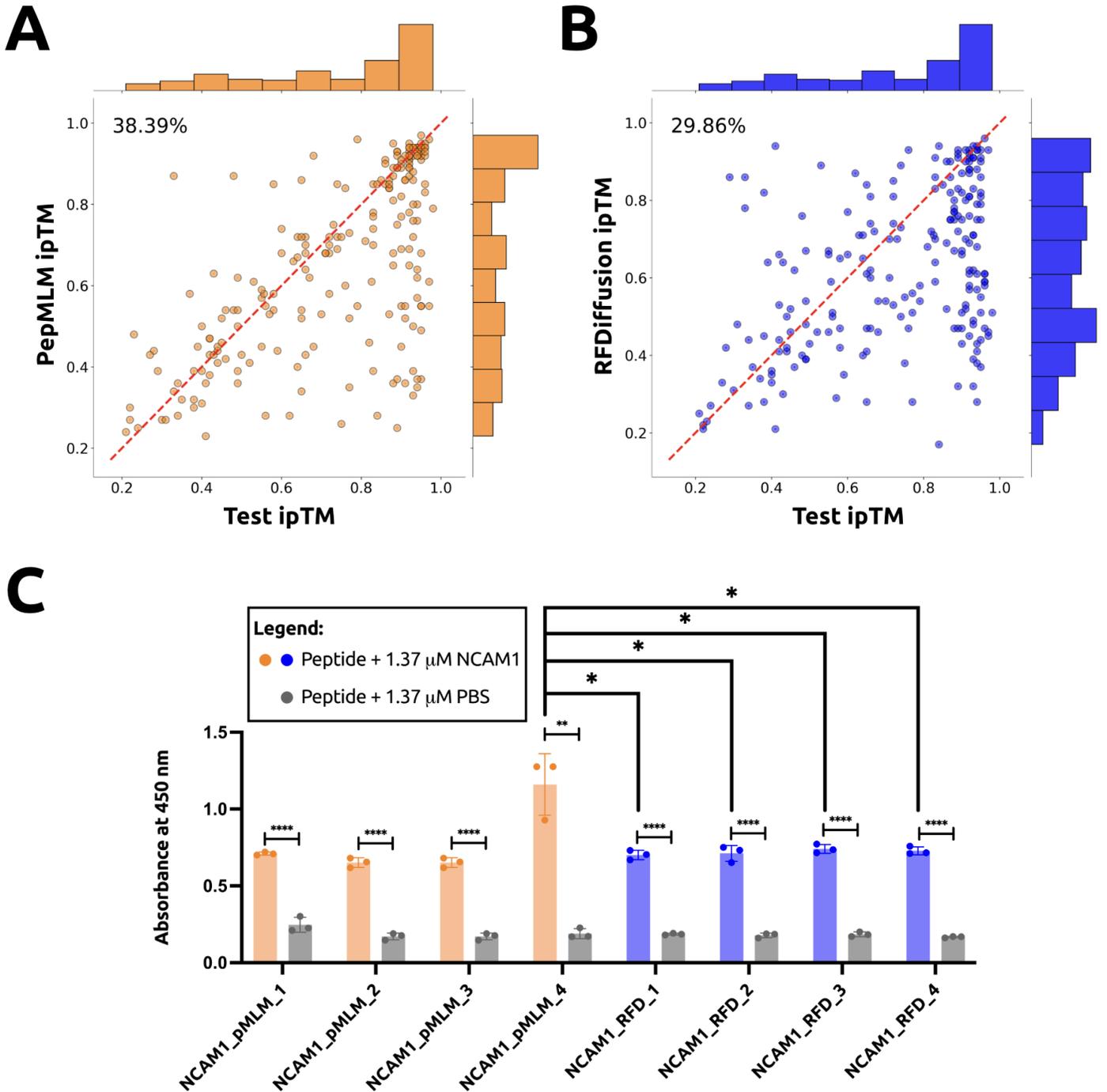

**Figure 2. Extrinsic evaluation of PepMLM-generated peptides *in silico* and *in vitro*.** (A) *In silico* hit-rate assessment of PepMLM. Utilizing AlphaFold-Multimer, ipTM scores were computed for both the generated and test peptides in conjunction with the target protein sequence. The entries are organized in accordance with the ipTM scores attributed to the test set peptides. The hit rate is characterized by the generated peptides exhibiting ipTM scores ≥ those of the test peptides. (B) *In silico* hit-rate assessment of RFDiffusion. The analogous assessment was applied for binders generated by RFDiffusion as for PepMLM-derived binders in Part A. (C) ELISA binding data of tested peptides at 1.37 uM of either NCAM1 ECD or PBS. Absorbance was calculated at 450 nm.

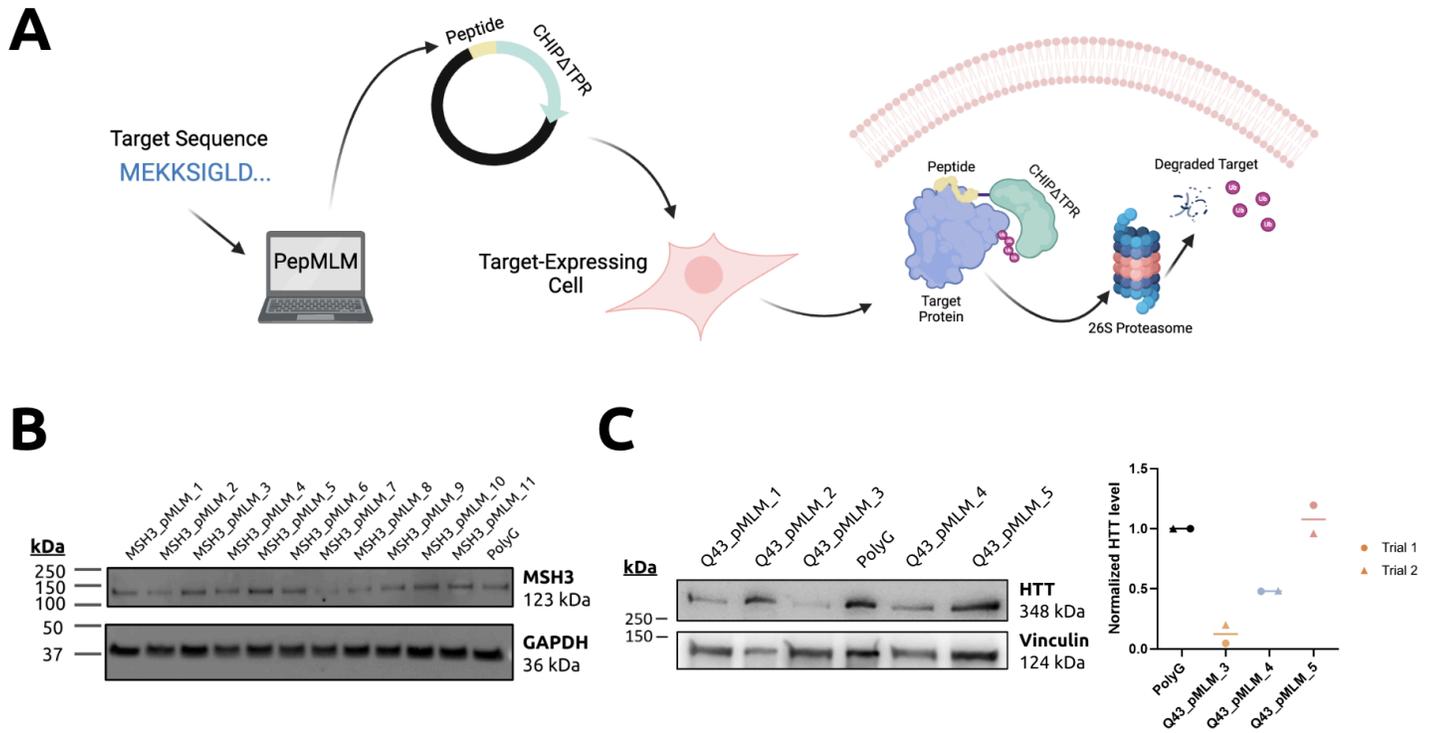

**Figure 3. Degradation of Huntington's disease-driving proteins *in vitro* with PepMLM-derived uAbs.** (A) Architecture and mechanism of uAb degradation system. CHIPΔTPR is fused to the C-terminus of PepMLM-designed target-specific peptides, and can thus tag endogenous target proteins for ubiquitin-mediated degradation in the proteasome, post-plasmid transfection. (B) Western blot analysis of HEK293T cells transfected with PepMLM peptides (MSH3_pMLM_1-11). (C) Western blot analysis of TruHD-Q43Q17M cells expressing PepMLM peptides (Q43_pMLM_1-5). Due to slight variance in the loading control levels, signal intensities for Q43_pMLM_3-5 were quantified using ImageJ and plotted using GraphPad, normalized to the PolyG control.

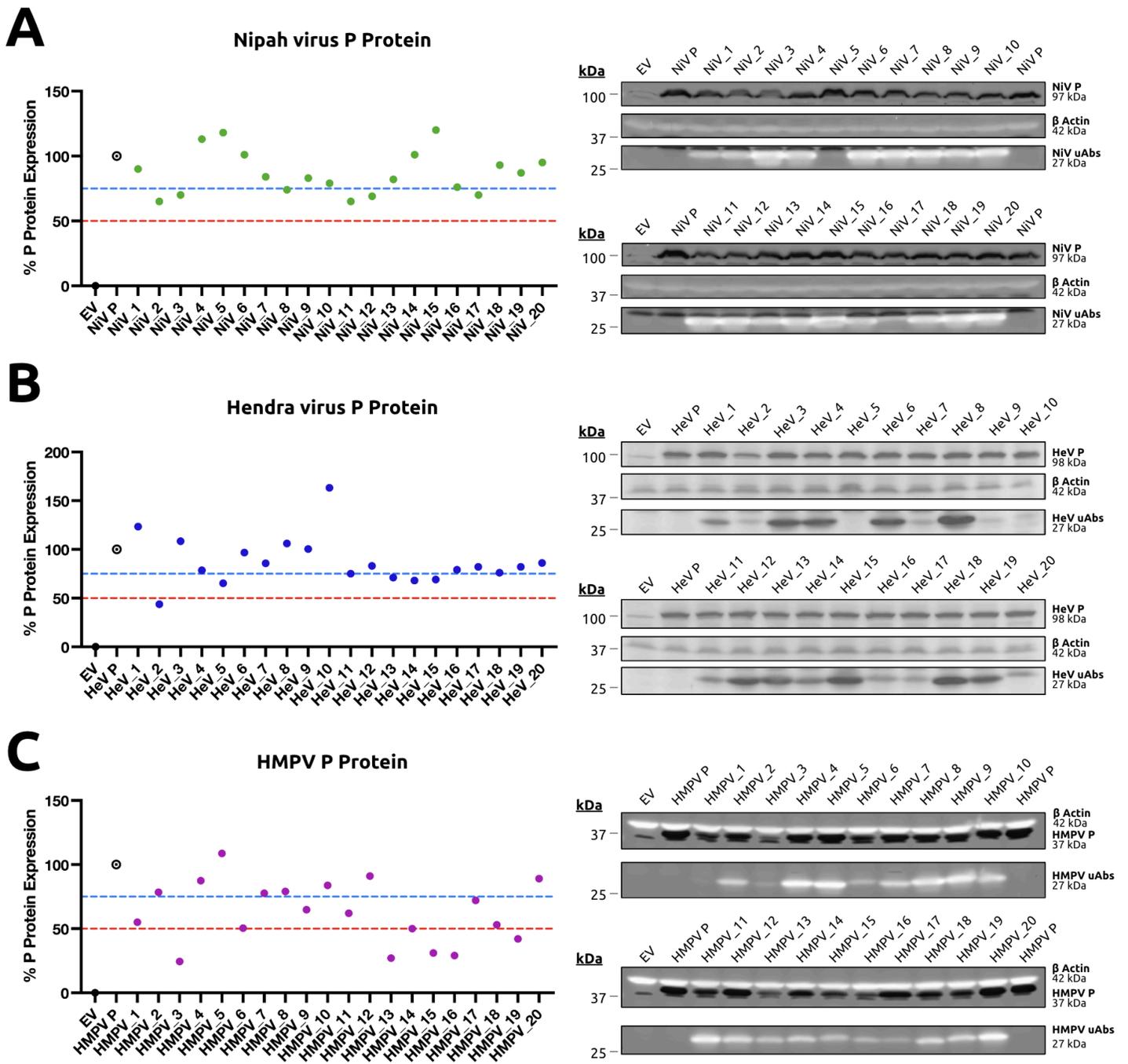

Figure 4. Screening of antiviral PepMLM-derived uAbs *in vitro*. 20 uAb plasmids were co-transfected with plasmid DNA for each of the phosphoproteins from (A) Nipah virus, (B) Hendra virus, and (C) Human metapneumovirus (HMPV) in HEK293T cells using PEI-Max. Whole cell lysates were harvested 48 hours post-transfection using RIPA buffer according to the manufacturer's protocol. uAbs and phosphoproteins were probed using mouse anti-FLAG and rabbit anti-HA antibodies, respectively, in addition to a mouse anti-β-actin loading control antibody. EV is an empty pCAGGS vector and P is a phosphoprotein-only control. Quantification of reduced detection of target P protein was determined by densitometry as described in Materials and Methods.

**PepMLM: Target Sequence-Conditioned Generation of Therapeutic Peptide Binders via Span Masked Language Modeling**

Tianlai Chen,[1] Madeleine Dumas,[2,3] Rio Watson,[1] Sophia Vincoff,[1] Christina Peng,[4] Lin Zhao,[1] Lauren Hong,[1] Sarah Pertsemlidis,[1] Mayumi Shaepers-Cheu,[2] Tian Zi Wang,[1] Divya Srijay,[1] Connor Monticello,[4] Pranay Vure,[1] Rishab Pulugurta,[1] Kseniia Kholina,[1] Shrey Goel,[1] Matthew P. DeLisa,[5-7] Ray Truant,[4] Hector C. Aguilar,[2] Pranam Chatterjee[1,8,9,†]

1. Department of Biomedical Engineering, Duke University
2. Department of Microbiology and Immunology, College of Veterinary Medicine, Cornell University
3. Department of Microbiology, College of Agriculture and Life Sciences, Cornell University
4. Department of Biochemistry and Biomedical Sciences, McMaster University
5. Meinig School of Biomedical Engineering, Cornell University, Ithaca, NY, USA
6. Robert F. Smith School of Chemical and Biomolecular Engineering, Cornell University, Ithaca, NY, USA
7. Cornell Institute of Biotechnology, Cornell University, Ithaca, NY, USA
8. Department of Computer Science, Duke University
9. Department of Biostatistics and Bioinformatics, Duke University

[†]Corresponding author: pranam.chatterjee@duke.edu

**Supplementary Figures**

1. Top-*k* sampling and perplexity.
2. Association between model perplexity and co-folding metrics.
3. Visualization of selected binder-protein complexes
4. Evaluation of PepMLM-3B.

**Supplementary Tables**

1. Settings and hyperparameters used to train PepMLM-650M.
2. Outlier analysis of protein-peptide complexes.
3. Selected sequence information and folding metrics for three classes.
4. Peptide sequences and PPL scores for experimental testing.

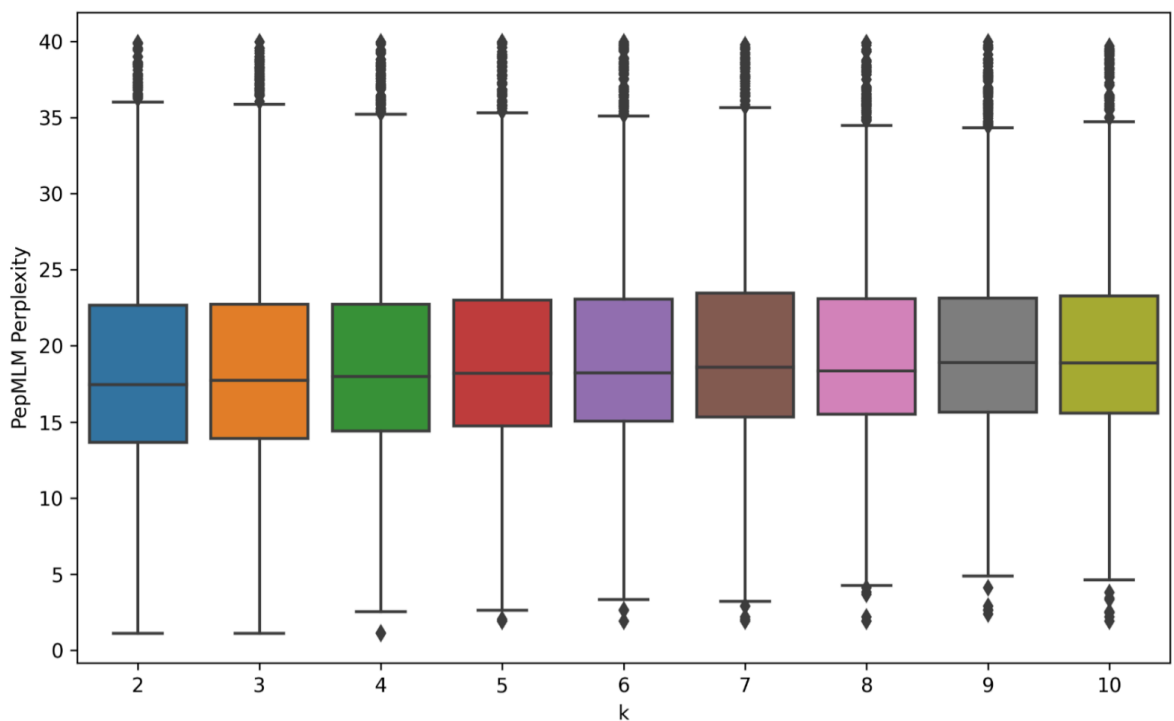

**Supplementary Figure 1. Top-*k* sampling and perplexity.** This figure illustrates the correlation between selected k values and the resulting Perplexity when generating binder sequences for target proteins, where each k value corresponds to the generation of 10 binders with lengths equating to the reference binder. As k escalates, there is a corresponding increase in Perplexity, indicating reduced model confidence. At k=3, the model maintains a lower Perplexity, signifying higher reliability in its predictions. The choice of k=3 is supported not only by this enhanced assurance but also by considerations of diminishing returns; beyond this point, the gain in diversity is outweighed by the increase in complexity and the potential for atypical predictions. Moreover, practical considerations favor k=3, as it offers computational efficiency without compromising the diversity necessary for effective binder design. Consequently, k=3 is endorsed as the optimal value to achieve a balance between model confidence, diversity of binder sequences, and computational pragmatism.

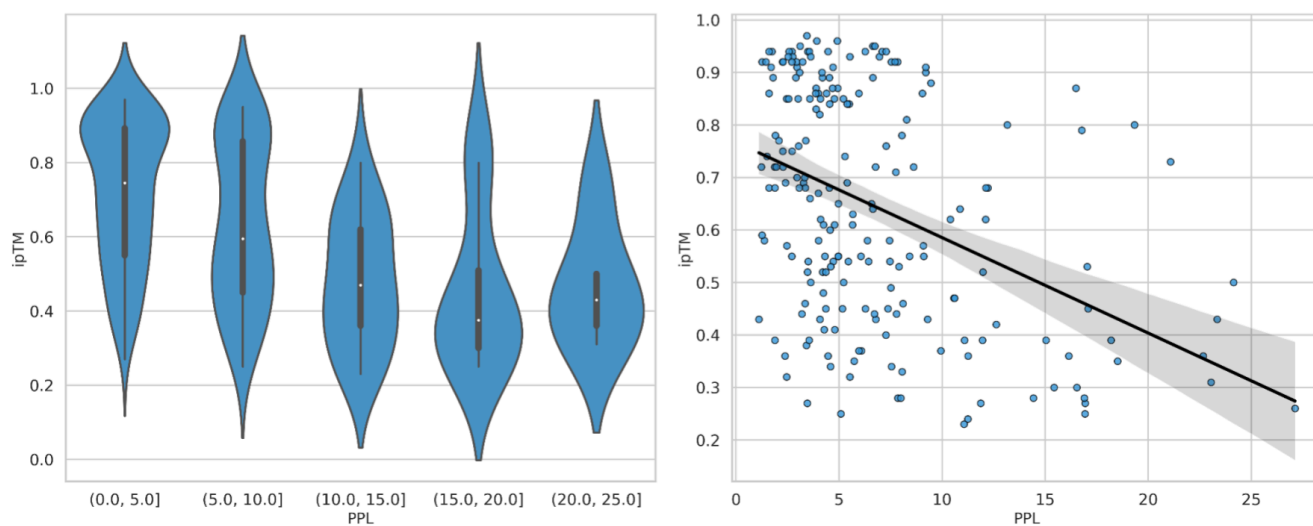
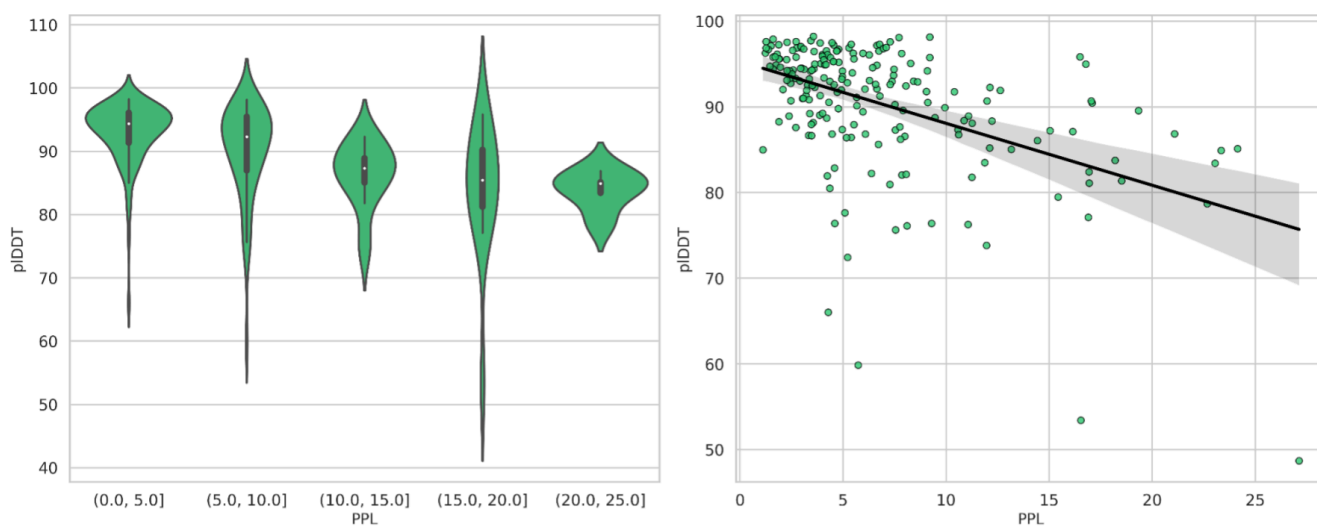

**Supplementary Figure 2. Association between model perplexity and co-folding metrics.** (A) Relationship between ipTM and perplexity (PPL). The initial segment of Figure A presents a violin plot, categorizing perplexity in 5-unit intervals. The subsequent segment delineates the raw data points, accompanied by a regression analysis, indicating a negative correlation (Pearson correlation coefficient -0.414, $p < 0.001$). The shaded area represents a 95% confidence interval. (B) Negative correlation between PPL and pLDDT, identified by Pearson correlation coefficient of -0.490 ($p < 0.001$). The violin plot underscores a marked decrement in specific folding metrics, most pronounced in ipTM, commensurate with elevated perplexity levels.

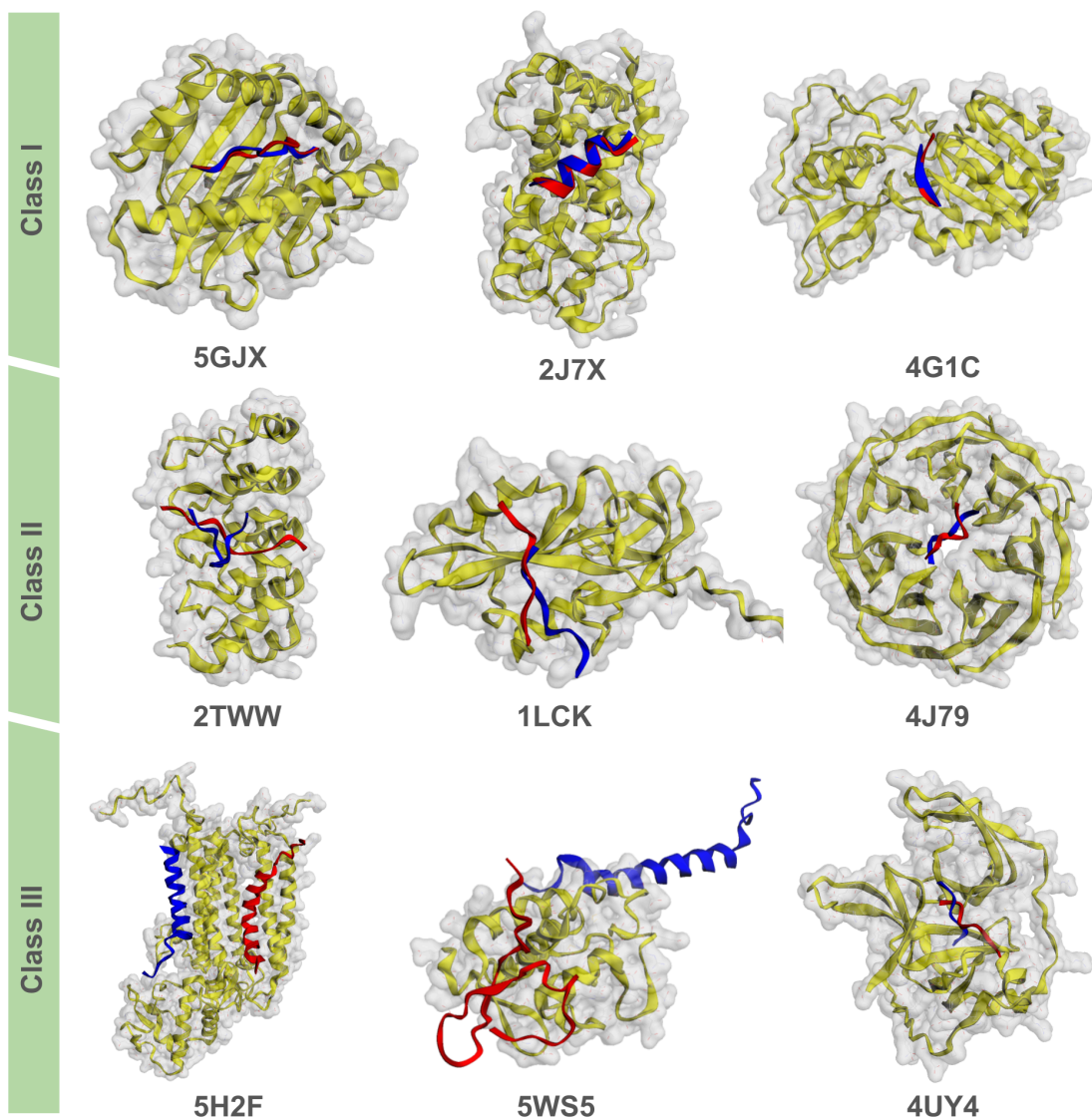

**Supplementary Figure 3. Visualization of binder-protein complexes.** Co-folded binder-protein complexes are categorized into three distinct classes for visualization purposes. Class I includes complexes where both the generated and test binders exhibit ipTM scores ≥ 0.7, Class II encompasses those with generated binders having ipTM scores ≥ 0.7 and test binders with ipTM scores < 0.7, and Class III contains complexes with both generated and test binders having ipTM scores ≤ 0.7. In these representations, the target protein is depicted in yellow, while the PepMLM-generated binders and test binders are illustrated in red and blue, respectively. This classification facilitates a detailed comparison of the structural relationships and binding patterns among the different classes of binder-protein complexes.

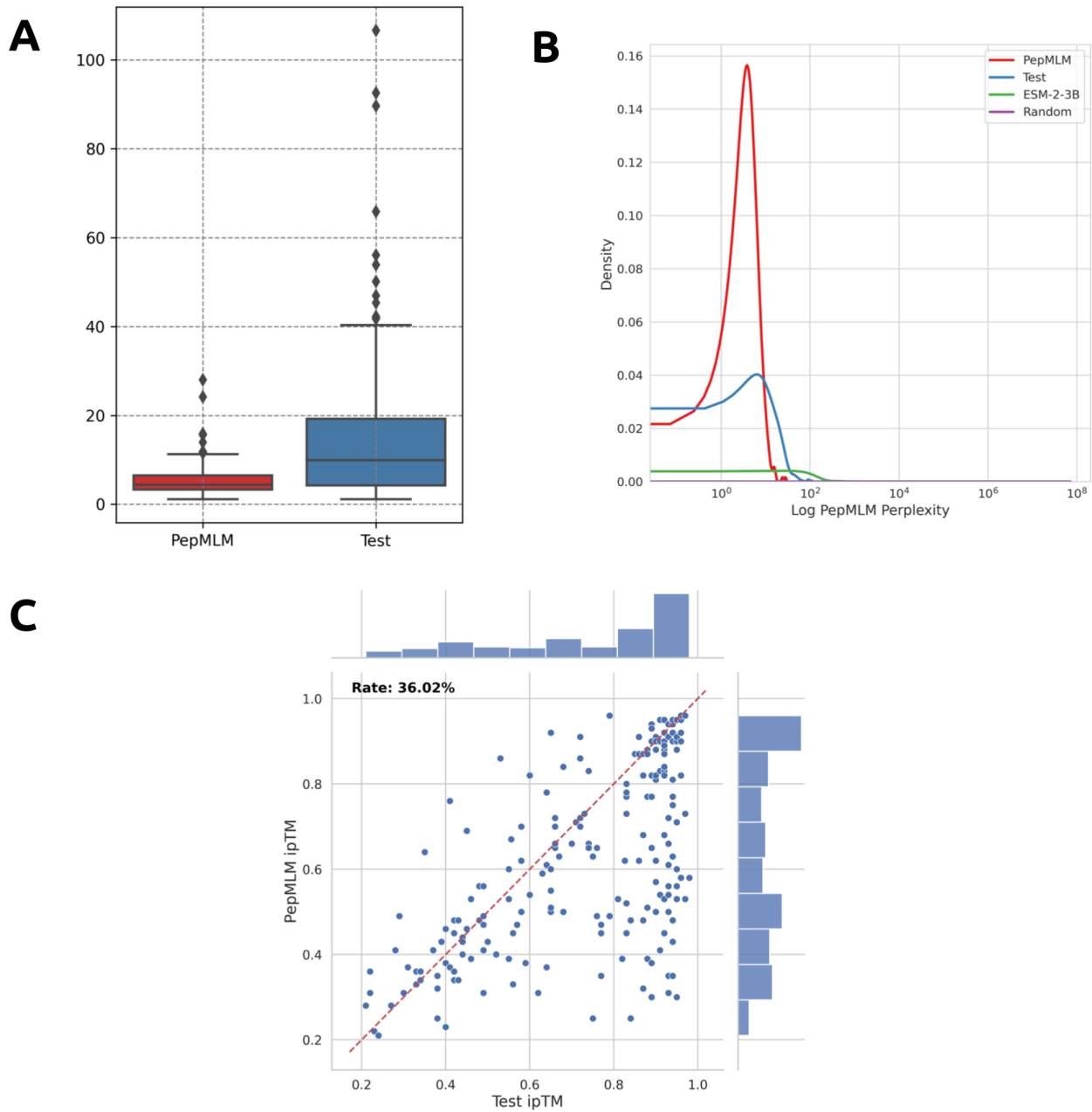

**Supplementary Figure 4. Evaluation of PepMLM-3B.** (A) Perplexity distribution comparison. The perplexity values were calculated for test and generated peptides, encompassing the target proteins in the test set. (B) The density distribution visualization of the log perplexity values for target-peptide pairs, encompassing test peptides, PepMLM-3B-generated peptides, ESM-2-3B-generated peptides, and random peptides. (C) *In silico* hit-rate assessment. Utilizing AlphaFold-Multimer, the ipTM scores were computed for both the generated and test peptides in conjunction with the target protein sequence. The entries are organized in accordance with the ipTM scores attributed to the test set peptides. The hit rate is characterized by the generated peptides exhibiting ipTM scores ≥ those of the test set peptides.

**Supplementary Table 1. Settings and hyperparameters used to train PepMLM-650M.**

| Name | Setting |
|---|---|
| Learning Rate | 0.000798 |
| Batch Size | 16 |
| Gradient Accumulation | 2 |
| Warm up steps | 501 |
| Number of training epochs | 5 |
| Optimizer | AdamW (default) |
| Trainer | Default HuggingFace Settings |

**Supplementary Table 2. Outlier analysis of protein-peptide complexes**. This table displays 12 protein-peptide complexes with pseudo-perplexity (PPL) values exceeding 40. Included are evaluation metrics for both the test complexes and the PepMLM generation results, as well as the binder sequences. ipTM scores for test and generated complexes are highlighted in different colors for comparison. Notably, even though these outliers exhibit high PPL values indicative of accurate modeling by PepMLM, the model remains proficient in generating binders that perform equivalently well *in silico* as per AlphaFold-Multimer ipTM score.

|        | Test | | | | PepMLM Generation | | | |
|--------|------|------|------|------|------|------|------|------|
| PDB ID | Binder | PPL | ipTM | pLDDT | Binder | PPL | ipTM | pLDDT |
| 5B5V | FLFGSRSS | 42.8 | 0.45 | 88.9 | YHYVMRYA | 4.2 | 0.52 | 88.7 |
| 4G1C | AVXCAX | 82.5 | 0.86 | 97.0 | TAKXST | 3.0 | 0.91 | 96.9 |
| 2L1C | RAKWDTANNPLXKEATSTFTNITXRGT | 45.4 | 0.49 | 61.9 | HIAEEPHFFESMQNNYEKPTTYKFQQK | 12.0 | 0.39 | 73.8 |
| 6GHJ | FAQ | 209.7 | 0.71 | 93.4 | MXL | 3.4 | 0.68 | 93.3 |
| 6AMU | MMWDRGLGMM | 59.2 | 0.34 | 87.9 | YQALIGGFNA | 14.4 | 0.28 | 86.1 |
| 5WMR | QIKVRVDMV | 76.3 | 0.91 | 92.3 | LRFWRARTL | 9.0 | 0.86 | 91.8 |
| 5NJC | VLEDRI | 63.0 | 0.83 | 97.5 | AAAAAA | 1.5 | 0.74 | 97.1 |
| 5FML | LSNDISQGIKRQRMTVESM | 42.8 | 0.91 | 94.0 | AAMTKLALAAKTRAQLFKK | 16.1 | 0.36 | 87.1 |
| 6DQU | GIINTL | 65.8 | 0.87 | 97.7 | YLGANG | 5.4 | 0.84 | 97.3 |
| 2IWB | GHMS | 194.0 | 0.64 | 96.1 | XPPX | 4.0 | 0.67 | 95.9 |
| 4MLI | AHIVMVDAYKPT | 62.6 | 0.87 | 97.5 | GPTPVQVLKRRG | 17.0 | 0.53 | 90.7 |
| 5DHM | RSIEISIRVDDFTKTGETVRY | 64.2 | 0.93 | 94.4 | AQSPEIITADVVVTSDEFTTT | 19.3 | 0.8 | 89.6 |

**Supplementary Table 3. Sequence information and folding metrics for complexes in Supplementary Figure 3.**

| PDB ID | Generated Binder | pIDDT | ipTM | Test Binder | pIDDT | ipTM |
|---|---|---|---|---|---|---|
| 5GJX | RLLEWMIYI | 96.3 | 0.92 | RLIQNSITI | 96.0 | 0.92 |
| 2J7X | HHLLLHLLTQD | 91.9 | 0.92 | IQSLINLLADN | 91.9 | 0.91 |
| 4G1C | TAKXST | 96.9 | 0.91 | AVXCAX | 97.0 | 0.86 |
| 3TWW | RREPPGGAFRX | 97.4 | 0.87 | RQSPDGQSFRX | 92.7 | 0.48 |
| 1LCK | PPXEEIPP | 87.3 | 0.92 | EGQQPQPA | 86.1 | 0.68 |
| 4J79 | AARHLD | 97.3 | 0.72 | EKVHVQ | 97.2 | 0.64 |
| 5H2F | XETNTLVRYVVAHFVLLVSVILIREAPRIESSKXX | 84.9 | 0.43 | XETITYVFIFACIIALFFFAIFFREPPRITXXXXX | 86.7 | 0.27 |
| 5WS5 | SSEEGRPILWIATTTGGGGVIIIVLFLFYAYYGSLSXLXXX | 77.7 | 0.24 | MSEGGRIPLWIVATVAGMGVIVIVGLFFYGAYAGLGSSLXX | 86.4 | 0.24 |
| 4UY4 | ARTKQT | 90.1 | 0.63 | ARTXQT | 89.0 | 0.43 |

**Supplementary Table 4. Peptide sequences and PPL scores for experimental testing.** The accession IDs for NCAM1, MSH3, and HTT are from UniProt. Viral target sequences are derived from NCBI/GenBank accession IDs.

| Target | Accession ID | uAb Name | Peptide Sequence | Derivation Algorithm | PPL |
|---|---|---|---|---|---|
| NCAM1 | P13591 | NCAM1_pMLM_1 | GKLPLPSLPCK | PepMLM | 5.343684945 |
| NCAM1 | P13591 | NCAM1_pMLM_2 | GLGPSPVLPRC | PepMLM | 5.299030403 |
| NCAM1 | P13591 | NCAM1_pMLM_3 | GLGPLPVLPCK | PepMLM | 6.329989946 |
| NCAM1 | P13591 | NCAM1_pMLM_4 | HSLGQPLSPIC | PepMLM | 4.753515747 |
| NCAM1 | P13591 | NCAM1_RFD_1 | SLPIENIYIEA | RFDiffusion | N/A |
| NCAM1 | P13591 | NCAM1_RFD_2 | MKPIEVVYEKA | RFDiffusion | N/A |
| NCAM1 | P13591 | NCAM1_RFD_3 | ELPEQVIYIEA | RFDiffusion | N/A |
| NCAM1 | P13591 | NCAM1_RFD_4 | EKPIEVIYEKA | RFDiffusion | N/A |
| MSH3 | P20585 | MSH3_pMLM_1 | SRREQLARILEGAFLASK | PepMLM | 7.65 |
| MSH3 | P20585 | MSH3_pMLM_2 | SRLEESASAMEASAAQAS | PepMLM | 9.12 |
| MSH3 | P20585 | MSH3_pMLM_3 | SRLKQAKSIMGGSLLLAE | PepMLM | 9.61 |
| MSH3 | P20585 | MSH3_pMLM_4 | NRLVEALASLEFSAQLSE | PepMLM | 9.91 |
| MSH3 | P20585 | MSH3_pMLM_5 | SRNKELKSILEFSLAQQK | PepMLM | 11 |
| MSH3 | P20585 | MSH3_pMLM_6 | SRLKQLASALDGSFLQAS | PepMLM | 11.84 |
| MSH3 | P20585 | MSH3_pMLM_7 | SLRKELASAMEFAAAQSK | PepMLM | 12.24 |
| MSH3 | P20585 | MSH3_pMLM_8 | SLNEQAASILEAFFAQSS | PepMLM | 13.47 |
| MSH3 | P20585 | MSH3_pMLM_9 | SYNVELASISEASLAAAK | PepMLM | 13.69 |
| MSH3 | P20585 | MSH3_pMLM_10 | SLNEQLASIMGGSAQLAE | PepMLM | 14.11 |
| MSH3 | P20585 | MSH3_pMLM_11 | SRRVELLSILEFALAAAS | PepMLM | 14.45 |
| HTT (Q43) | P42858 | Q43_pMLM_1 | SAAPQLLGSGLALGK | PepMLM | **Q43:** 5.302052023<br>**Q17:** 21.74186374 |
| HTT (Q43) | P42858 | Q43_pMLM_2 | TAPQLSQASGLAGGK | PepMLM | **Q43:** 6.223211251<br>**Q17:** 22.05928384 |
| HTT (Q43) | P42858 | Q43_pMLM_3 | LAPQLLLLGLGGLAK | PepMLM | **Q43:** 6.122010055<br>**Q17:** 21.22001317 |
| HTT (Q43) | P42858 | Q43_pMLM_4 | SAPPQLAAAGGLLLA | PepMLM | **Q43:** 5.42110285<br>**Q17:** 19.667137 |
| HTT (Q43) | P42858 | Q43_pMLM_5 | SPPPQAAAGAALGAK | PepMLM | **Q43:** 6.196895246<br>**Q17:** 20.36823003 |

| Hendravirus P | MN062017.1 | HeV_1 | RLPVYLSLQG | PepMLM | 3.635173 |
|---|---|---|---|---|---|
| Hendravirus P | MN062017.1 | HeV_2 | HSPVHLSLLG | PepMLM | 4.296802 |
| Hendravirus P | MN062017.1 | HeV_3 | SRSVLHSLLQGR | PepMLM | 4.974103 |
| Hendravirus P | MN062017.1 | HeV_4 | HSSVLQSLFGG | PepMLM | 5.449664 |
| Hendravirus P | MN062017.1 | HeV_5 | SESLYLSLFKG | PepMLM | 6.702855 |
| Hendravirus P | MN062017.1 | HeV_6 | SMSRRRQLAKKLLLLAIKS | PepMLM | 6.707319 |
| Hendravirus P | MN062017.1 | HeV_7 | RQSLRQQLLLDLGR | PepMLM | 6.74773 |
| Hendravirus P | MN062017.1 | HeV_8 | SSLVYLSLGA | PepMLM | 6.824329 |
| Hendravirus P | MN062017.1 | HeV_9 | HLSLPHSLLQKR | PepMLM | 6.850688 |
| Hendravirus P | MN062017.1 | HeV_10 | SMSVEKSLSKKLGKKLIKS | PepMLM | 6.855796 |
| Hendravirus P | MN062017.1 | HeV_11 | RSLVKKQLLLKLLG | PepMLM | 6.950253 |
| Hendravirus P | MN062017.1 | HeV_12 | MQSVKLKLLLKGLLR | PepMLM | 7.039044 |
| Hendravirus P | MN062017.1 | HeV_13 | HSSVLQQLFGE | PepMLM | 7.046471 |
| Hendravirus P | MN062017.1 | HeV_14 | HHSLLQSLLQGT | PepMLM | 7.287729 |
| Hendravirus P | MN062017.1 | HeV_15 | RLPLYLSLGA | PepMLM | 7.372559 |
| Hendravirus P | MN062017.1 | HeV_16 | HHSLLHSLLKGT | PepMLM | 7.621191 |
| Hendravirus P | MN062017.1 | HeV_17 | RLSVLQQLLKLGG | PepMLM | 7.758791 |
| Hendravirus P | MN062017.1 | HeV_18 | SLSRRQQLLLDLGK | PepMLM | 7.797618 |
| Hendravirus P | MN062017.1 | HeV_19 | SKPLYLLLGG | PepMLM | 7.82979 |
| Hendravirus P | MN062017.1 | HeV_20 | RLSVRKLLLLDLGK | PepMLM | 7.860104 |
| HMPV P | AAS22075.1 | HMPV_1 | LTLEQLQEKR | PepMLM | 6.510858 |
| HMPV P | AAS22075.1 | HMPV_2 | TLEEELLLKR | PepMLM | 7.514123 |
| HMPV P | AAS22075.1 | HMPV_3 | LTLEQLQLIR | PepMLM | 7.855078 |
| HMPV P | AAS22075.1 | HMPV_4 | AELLLRQQQLLL | PepMLM | 7.869746 |
| HMPV P | AAS22075.1 | HMPV_5 | SVLTAEQLIKI | PepMLM | 8.039237 |
| HMPV P | AAS22075.1 | HMPV_6 | DLRRRLAEKTPELQLLLI | PepMLM | 8.068756 |
| HMPV P | AAS22075.1 | HMPV_7 | ALLAKKLTLEALLAL | PepMLM | 8.235918 |
| HMPV P | AAS22075.1 | HMPV_8 | AEEAKKLTEELLRLR | PepMLM | 8.411599 |
| HMPV P | AAS22075.1 | HMPV_9 | DTELAAKKLTTELLLKI | PepMLM | 8.626335 |
| HMPV P | AAS22075.1 | HMPV_10 | TLTLQQLLKL | PepMLM | 8.713968 |